# Measurement and assignment of $J \geq 10$ rotational energy levels in the 9510 to 9810 $cm^{-1}$ and 6590 to 6900 $cm^{-1}$ ranges of methane using optical frequency comb double-resonance spectroscopy

**Yuan Cao[1], Adrian Hjältén[1], Vinicius Silva de Oliveira[1], Isak Silander[1], Michael Rey[2], Kevin K. Lehmann[3], and Aleksandra Foltynowicz[1*]**

[1] Department of Physics, Umeå University, 901 87 Umeå, Sweden

[2] Laboratoire Interdisciplinaire Carnot de Bourgogne, UMR CNRS 6303, Université Bourgogne Europe, 9 Av. A. Savary, BP 47870, 21078 Dijon Cedex, France.

[3] Departments of Chemistry & Physics, University of Virginia, Charlottesville, VA 22904, USA

*aleksandra.foltynowicz@umu.se

**Abstract**: Accurate models of high temperature methane spectra are needed in astrophysics. Previous measurements of methane hot-band transitions in the $P6 \leftarrow P2$ polyad range have been limited to final rotational numbers of $J \leq 9$, with theoretical predictions at higher $J$ remaining unvalidated. Here, we use optical-optical double resonance spectroscopy (OODR) with a 3.3 µm narrow linewidth pump to excite the $\nu_3$ P(12, $A_1^{(2)}$) methane transition ($P2 \leftarrow P0$) and a cavity-enhanced frequency comb centered around 1.68 µm to probe the sub-Doppler ladder-type ($P6 \leftarrow P2$) and V-type ($P4 \leftarrow P0$) transitions, as well as Doppler-broadened collision-induced four-level transitions ($P6 \leftarrow P2$). 49 ladder-type transitions with final rotational states $J$ = 10-12 in the range of 9510 to 9810 $cm^{-1}$ (i.e., the $P6$ polyad) were assigned to effective Hamiltonian predictions and the ExoMol database, of which 6 reached vibrational states that had not been observed experimentally before. 19 sub-Doppler V-type transitions with final states $J$ = 11-13 in the range of 6590 to 6900 $cm^{-1}$ (i.e., the $P4$ polyad) were observed and assigned to the Hamiltonian and ExoMol, while only 2 of these V-type transitions could be unambiguously assigned to WKLMC and HITRAN line lists. 170 Doppler-broadened four-level double-resonance (4LDR) lines were observed, 7 of which were newly observed compared with our previous work when pumping transitions starting from the $J$ = 7 level in the ground state [Lehmann *et al*., J. Chem. Phys. **163**, 144304 (2025)]. We could not assign these lines as they did not form combination differences with other observed 4LDR transitions.

## 1. Introduction

Methane ($CH_4$) is an important research subject in molecular spectroscopy due to its crucial role in analyzing the atmospheres of Earth and exoplanets [1]. Despite its simple tetrahedral symmetry, $CH_4$ displays highly complex rovibrational spectra [2]. Spectral models used to predict $CH_4$ spectra in high temperature environments employ extensive lists of spectral transitions, including many hot-bands [3, 4]. Hot-band transition measurements of $CH_4$ are hence essential for verifying and improving the accuracy of such models, which in turn can be used for the simulation of the high temperature spectra [3, 5, 6].

The vibrational energy levels of $CH_4$ form clusters of strongly interacting states, known as polyads, due to the near coincidence of fundamental and overtone frequencies and the strong near-resonant couplings (Fermi, Coriolis, Darling-Dennison) between them. The polyad number $P$ is defined based on the principal vibrational normal mode quantum numbers ($\nu_i$): $P=2\nu_1+\nu_2+2\nu_3+\nu_4$, which groups levels with similar energies [7]. Algebraic Hamiltonian models including these intra-polyad couplings have allowed for accurate spectroscopic modeling [8]. However, the number of sublevels within each polyad increases exponentially with the polyad number, leading to increased spectral complexity [9].

Consequently, firm state assignments are scarce for states in *P*6 polyad and become even more limited for higher polyads.

While laser-based spectroscopy has been used to characterize the $CH_4$ line lists [10-18], most line lists above 8000 $cm^{-1}$ are primarily derived from low- and room-temperature Fourier transform infrared (FTIR) spectroscopy [19-22]. Such measurements cannot provide information of hot-band transitions starting from states that are not populated at room temperature.

The ExoMol database provides comprehensive high-temperature (up to 2000 K) $CH_4$ line lists, primarily used for exoplanet and stellar atmosphere modeling [23]. The latest ExoMol line list contains over 50 billion transitions between over 9 million energy levels and covers wavenumbers up to 12000 $cm^{-1}$ with total angular momentum $J \leq 60$ [23]. It was generated through solution of the nuclear motion Schrödinger equation for an empirically derived potential energy surface (PES) and a new high-level *ab initio* dipole moment surface (DMS). The theoretical predictions in ExoMol had been augmented by replacing calculated term values by experimentally derived term values from the literature using the MARVEL (Measured Active Rotational Vibrational Energy Levels) algorithm, which has led to the incorporation of 23208 empirical term values, including uncertainties, with $J \leq 27$ below 9986 $cm^{-1}$, covering the lowest eight polyads [9, 23]. However, the MARVEL list remains scarce for high energy and high-*J* term values due to the limited availability of assigned spectra involving such states. Recently, a new methodology was proposed by Rey [8] for the construction of *ab initio* effective Hamiltonian model from PES and DMS. ExoMol and Hamiltonian are based on two different methodologies. In ExoMol, huge matrices (>100000) are diagonalized, including inter-polyad couplings. However, the Hamiltonian involves much smaller matrices (~1000), which are restricted to a given polyad (the so-called effective models); interpolyad couplings are removed through successive numerical block diagonalizations, as an alternative to contact transformations. Comparison of both these prediction sets to measured transitions in the $P6 \leftarrow P2$ range of $CH_4$ for rotational numbers $J \leq 9$, has shown that the Hamiltonian predictions are more accurate than those of ExoMol [24, 25].

Optical-optical double resonance (OODR) spectroscopy is a powerful tool for detecting hot-band transitions from one selected excited rovibrational state and determining the final upper state rotational assignments [26-30]. We have previously demonstrated OODR spectroscopy using a 3.3 μm continuous wave (CW) optical parametric oscillator (OPO) pump and a 1.68 μm frequency comb probe to record the $P6 \leftarrow P2$ and $P4 \leftarrow P0$ transitions in $CH_4$, which combined the advantages of sub-Doppler resolution and wide spectral coverage [31, 32]. Using this technique, we have until now detected 241 ladder-type ($P6 \leftarrow P2$) transitions reaching final states in the range of 8580-9370 $cm^{-1}$ with rotational numbers $J \leq 9$, and 56 V-type ($P4 \leftarrow P0$) transitions reaching final states in the 5600-6360 $cm^{-1}$ region with $J \leq 8$ [24, 31-34].

The aim of this work is to further expand the *J*-range of experimentally observed term values and compare against the theory. We use OODR spectroscopy to measure overtone and hot-band transitions of $CH_4$ in the 5741-6120 $cm^{-1}$ probe range reaching states with rotational quantum number $J \geq 10$. Fig. 1 schematically illustrates the vibrational bands and transitions addressed by the CW pump (green) and comb probe (red) in this work. In Fig. 1a), the CW OPO laser pumps a narrow velocity distribution of molecules from the ground state to the selected rotational level in the *P*2 polyad, and then a cavity-enhanced frequency comb excites the sub-Doppler ladder-type transitions from the pumped level to the *P*6 region. In addition, the comb also excites Doppler-broadened (DB) transitions from thermally populated levels in the ground state to the *P*4 region. When those DB transitions share the same lower state as the pump transition, sub-Doppler V-type transitions appear at the centers of the DB lines due to the pump-induced depletion of the lower state population. Fig. 1b) illustrates the rotational structure of the ladder-type transitions investigated in this work. We pumped the $\nu_3$ P(12, $A_1^{(2)}$) transition, and probe the transitions from the pumped state with $J' = 11$ to final states in *P*6 polyad with $J = 10$-12. Fig. 1c) shows the V-type transition from the pump-depleted state with $J''=12$ to states in the *P*4 polyad with *J* = 11-13. The population in the pumped *J'* = 11 level is also redistributed to other levels within the *P*2 polyad via inelastic collisions, leading to Doppler-broadened double-resonance hot-band transitions in the probe spectrum. These are referred to as four-level double-resonance (4LDR) transitions, as the pump and probe transitions do not share a common state, in contrast to the three-level double resonance (3LDR) ladder-type and V-type transitions [35]. Fig. 1d) shows the 4LDR transitions from collisionally populated excited states in the *P*2 polyad to final states in the *P*6 polyad.

We report 49 ladder-type and 19 V-type lines, and assign them to transitions in the effective Hamiltonian [8] and ExoMol [23] lists, respectively. Among these, 6 vibrational states in the *P*6 polyad have not been previously observed experimentally. The term values for the upper states observed in this work expand the range of experimentally verified rovibrational levels in both the *P*6 and *P*4 polyad regions.

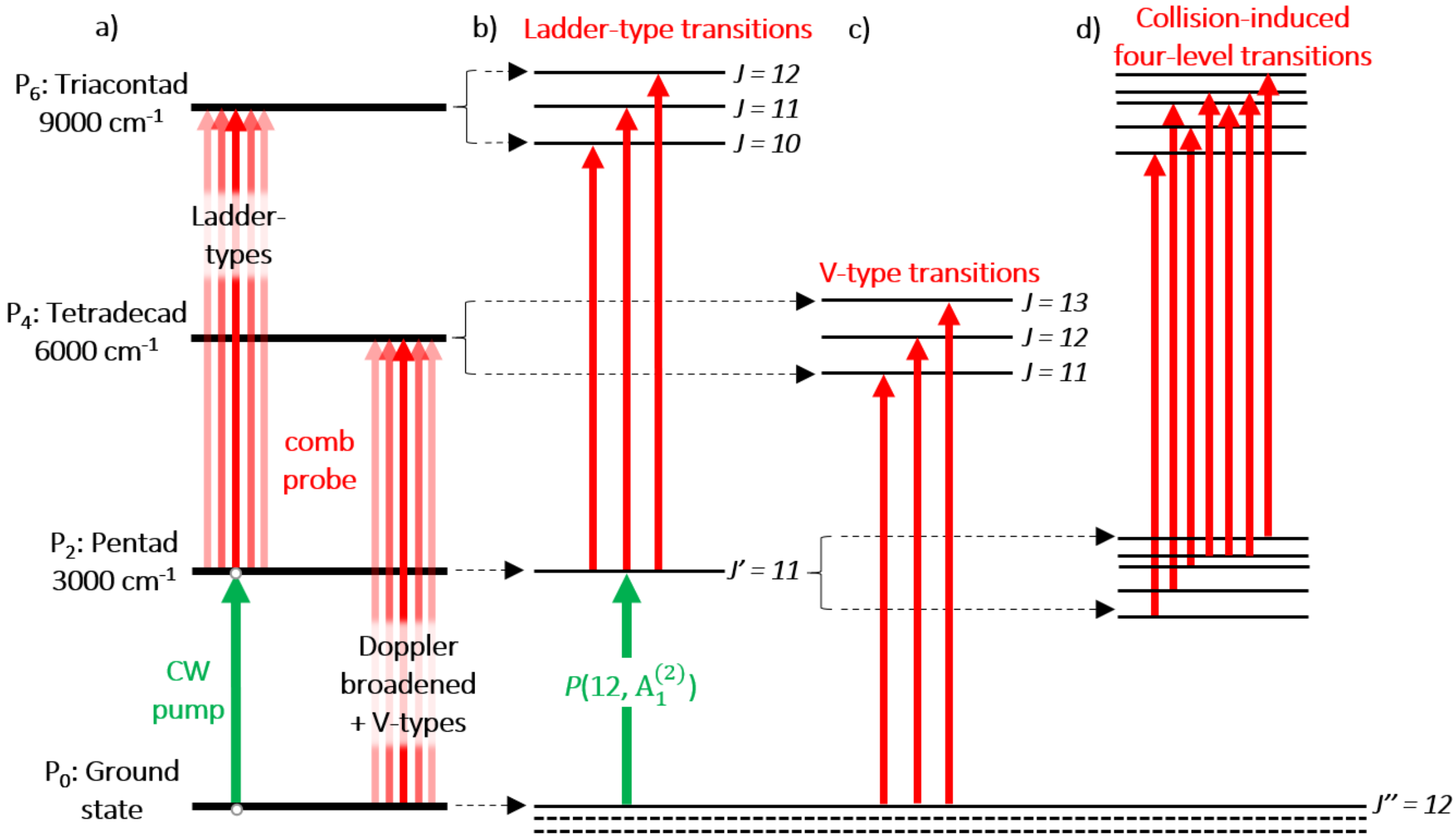

Fig. 1. a) Vibrational levels addressed by the CW pump (green) and comb probe (red). b) Detailed schematic of the rotational levels involving the ladder-type transitions when pumping the $\nu_3$ P(12, $A_1^{(2)}$) state. c) Detailed schematic of rotational levels involving the V-type transitions when pump and probe starting from the same *J''*=(12, $A_1^{(2)}$) state. d) Schematic depiction of the collision-induced four-level transitions.

## 2. Experimental setup and procedures

The OODR experimental setup is similar to that in the Ref. [34] and is only briefly summarized here. The OODR experimental setup employed a 3.3 μm CW OPO pump laser and a 1.68 μm cavity-enhanced frequency comb probe, with the probe spectrum analyzed by a Fourier transform spectrometer (FTS). The pump was the idler of a singly-resonant CW OPO (TOPTICA, TOPO) that was phase locked to a tooth of a fully stabilized 125 MHz mid-infrared (MIR) comb produced by difference frequency generation [36]. The probe was a 250 MHz Er:fiber optical frequency comb (Menlo Systems FC-1500-250-WG) that was frequency shifted by a polarization-maintaining microstructured silica fiber (PM-MSF). The center wavenumber of the probe can be tuned between 1.6 and 1.8 μm by adjusting the comb power coupled into the PM-MSF. The pump and probe beams were colinearly coupled into a 60-cm-long cavity. The comb beam was mode matched to the $TEM_{00}$ mode of the cavity with a finesse around 1100 and a free spectral range (FSR) matched to comb repetition rate, while the pump passed once through the cavity, overlapping the probe mode with the same focal point and confocal length. Dichroic mirrors were used to combine the pump and probe beams in front of the cavity and separate them at the output. The comb was locked to the cavity at a spectral point using the Pound-Drever-Hall (PDH) method [34], with feedback applied to $f_{rep}$ by means of a piezoelectric (PZT) actuator and the electro-optic modulator (EOM) within the comb oscillator cavity. To achieve absolute stabilization of $f_{rep}$, it was locked to a radio frequency (RF) source by feeding back to a PZT controlling the sample cavity length. The comb carrier-envelope-offset ($f_{ceo}$) was locked to another RF source, and its value was adjusted to maximize the spectral bandwidth of the cavity transmission.

The probe light transmitted through the cavity was sent to an FTS, and both FTS outputs were detected with a home-built auto-balanced InGaAs detector [37]. The optical path difference was calibrated with a frequency-stabilized HeNe laser (Sios, SL/02/1) operating around 632.991 nm (reference laser wavelength, $\lambda_{ref}$). By utilizing the sub-nominal resolution method [38, 39], the sampling points in the Fourier spectrum were matched to the comb tooth frequencies, such that each FTS scan yielded a spectrum with frequency points separated by $f_{rep}$. In order to observe the sub-Doppler OODR features, spectra were recorded while scanning $f_{rep}$ up and down in 130 steps with a step size of 2.75 Hz, corresponding to ~2 MHz steps in the optical domain. To minimize the impact of the background spectral features, a background interferogram was acquired at each $f_{rep}$ step with the pump laser blocked by a shutter. Multiple back-and-forth $f_{rep}$ scans were performed for spectral averaging. The relative polarization of the pump and probe beams was set to the magic angle (54.7 °) to remove the effect of pump-induced angular momentum alignment on the OODR probe absorption strength [40]. All fixed and tunable frequency sources and counters are referenced to a GPS-disciplined rubidium clock.

We selected the $\nu_3$ P(12, $A_1^{(2)}$) [$(P,J,C,N)$=(2,11,$A_2$,15)←(0,12,$A_1$,2)] transition for pumping because it is well isolated from neighboring transitions while still possessing relatively high line intensity. Here $P$ is the polyad number, $J$ is total angular momentum quantum number, $C$ is the rovibrational symmetry, and $N$ is the counting number distinguishing states within a $(P,J,C)$-block. In addition, ExoMol predicts 11 ladder-type transitions from upper level of the pump transition to vibrational states without MARVELized sub-levels in the probe region of 5400-6100 cm$^{-1}$, compared to only 4 and 3 new upper vibrational states for the Q(12, $A_1^{(2)}$) and R(12, $A_1^{(2)}$) pump transitions, respectively. The pump laser was locked to the transition wavenumber 2894.99609799(7) cm$^{-1}$ reported by Abe *et al* [13].

To maximize the observation of unMARVELized upper vibrational states, two measurements were performed over different probe ranges: measurement #1 covering 5740-5910 cm$^{-1}$ and measurement #2 covering 5870-6120 cm$^{-1}$. The spectral coverage was tuned by adjusting the comb power coupled into the PM-MSF, the PDH locking point (i.e., the portion of the spectrum used for locking), and the $f_{ceo}$. The PDH locking points were 5833 cm$^{-1}$ and 6029 cm$^{-1}$, the $f_{ceo}$ were -104 MHz and -34 MHz, the numbers of $f_{rep}$ scans recorded for averaging were 24 and 32, respectively, while the sample pressure was 200 mTorr for both measurements. Unlike the comb teeth, the cavity resonances are not precisely equally spaced due to the dispersion from the cavity mirror coatings and the sample gas, leading to increasing comb-cavity offsets at comb teeth further away from the locking point. Due to the narrowing of the comb soliton output and the comb-cavity phase offset, the comb bandwidth after transmission through the cavity decreased from ~7 THz in measurement #2 to ~5 THz in measurement #1.

## 3. Spectral analysis

The comb spectra were processed in two different ways. To detect OODR transitions and to fit ladder-type and 4LDR transitions, the comb spectra acquired with pump excitation were normalized to the background spectra acquired with pump blocked. This normalization process largely removes the Doppler-broadened overtone transitions and the comb envelope shape. To fit V-type transitions and to assess the cavity enhancement factor, we used the cepstral method to remove the comb envelope of the spectra, simulating the cavity-enhanced Doppler-broadened $CH_4$ absorption using parameters from HITRAN2020 [41], as described in Refs. [24, 34] In both approaches, the remaining baselines were modelled as a combination of a fifth-order polynomial curve and sine terms accounting for etalon fringes, and were cancelled from the spectrum by division. For the spectra interleaved without background normalization, we performed the baseline removal in 8 segments as this improved the results. The resulting baseline-corrected spectra acquired at different $f_{rep}$ values were then interleaved to produce a spectrum with a ~2 MHz spectral point spacing. Finally, we used the peak finding routine described previously [24, 35] on the background-normalized spectra to determine the positions of the sub-Doppler 3LDR lines and the Doppler broadened 4LDR lines, respectively. The cavity enhancement factor was determined from the spectra without background normalization by fitting a cavity transmission model to the Doppler broadened lines [25, 42]. In the overlapping region of the two probe ranges, no common OODR transition were observed due to the low signal-to-noise ratio (SNR) on the wings of the comb envelopes.

### *3.1 Fitting of OODR transitions*

We identified 49 ladder-type lines between 5798 to 6101 cm$^{-1}$. We fitted the lines as in previous work by modeling them as a sum of a sub-Doppler Lorentzian component and a broader Gaussian background due to elastic collision-induced velocity redistribution, both sharing a common center frequency in the cavity transmission function [43]. For lines with a Gaussian component SNR greater than 5, the free fitting parameters of the cavity transmission model included the center frequency, the integrated absorption and linewidth of both the Lorentzian and Gaussian components, the comb-cavity phase offset, and a first- or third-order polynomial to account for the baseline. We determined the mean value of the integrated absorption ratio of the Gaussian component to the Lorentzian component and the Gaussian linewidth from those strong ladder-type lines, and then fixed these parameters when fitting lines with a Gaussian component SNR below 5. For most lines, a fitting range of ±450 MHz around the line center and a first-order polynomial background were adopted, but the fitting range and polynomial order were sometimes adjusted for lines with pronounced baseline structure. The model also included the ground state absorption and dispersion of the Doppler-broadened background, simulated using the parameters in HITRAN2020 [41]. Fig. 2a) shows the observed strongest ladder-type line at 6061.335288(5) cm$^{-1}$ (black), along with the fit (red) and Lorentzian (green) and Gaussian (blue) fit components. The asymmetry of the line was caused by the non-zero comb-cavity phase offset, which was included in the model. The lower panel shows the fit residuals.

We also identified 19 V-type lines in the range 5775 to 6081 cm$^{-1}$. The fitting of these lines was performed as in Ref. [24] in the unnormalized spectra to preserve the Doppler-broadened features and avoid additional noise introduced by division with the background spectra. We fitted the Doppler-broadened line exhibiting the V-type feature as a Gaussian line shape within a fitting range of ±1 GHz around the line center. The fitting parameters included the center frequency, the integrated absorption of the Doppler component, and the comb-cavity phase offset. The V-type feature was then isolated by dividing the data by the fitted Gaussian profile, and subsequently fitted in a manner similar to the fitting of ladder-type transitions, but with the signs of both the Lorentzian and Gaussian components inverted. The integrated absorption ratio between the Lorentzian and Gaussian components, as well as the Gaussian line width, were fixed to values obtained from the fitting of the ladder-type transitions with a Gaussian component SNR greater than 5. For all V-type lines, the fitting window was set to ± 200 MHz, and the baseline was modeled with a third order polynomial. Fig. 2b) displays the V-type line at 5994.99534(1) cm$^{-1}$ (black), together with the fit (red) and Lorentzian (green) and Gaussian (blue) fit components. The lower panel shows the fit residuals.

We fitted 170 collision-induced 4LDR lines in the range of 5785 to 6076 cm$^{-1}$ using a cavity transmission Gaussian line shape model, with the fitting parameters including the center frequency, the integrated absorption coefficient, the Doppler-broadened linewidth, and the comb-cavity phase offset. The following fitting ranges and background treatments were applied sequentially until the fitting converged [35]: (1) a fitting range of ±1 GHz around each identified line, without including background; (2) reduction of the fitting range to ±400 MHz to avoid the baseline problem, still without background; (3) restoration of the ±1 GHz fitting range with the inclusion of a background offset; (4) the asymmetric fitting range of ±200 MHz to ±1 GHz was set for lines partially overlapping with neighboring features, again including a background offset; (5) restoration of the ±1 GHz or ±400 MHz fitting range, with the baseline modeled by a first-order polynomial. If the fit failed to converge after these steps, it was discarded. Fig. 2c) shows the measured 4LDR line at the 5818.877(1) $cm^{-1}$ (black) and the fit (red). The lower panel shows the fit residuals.

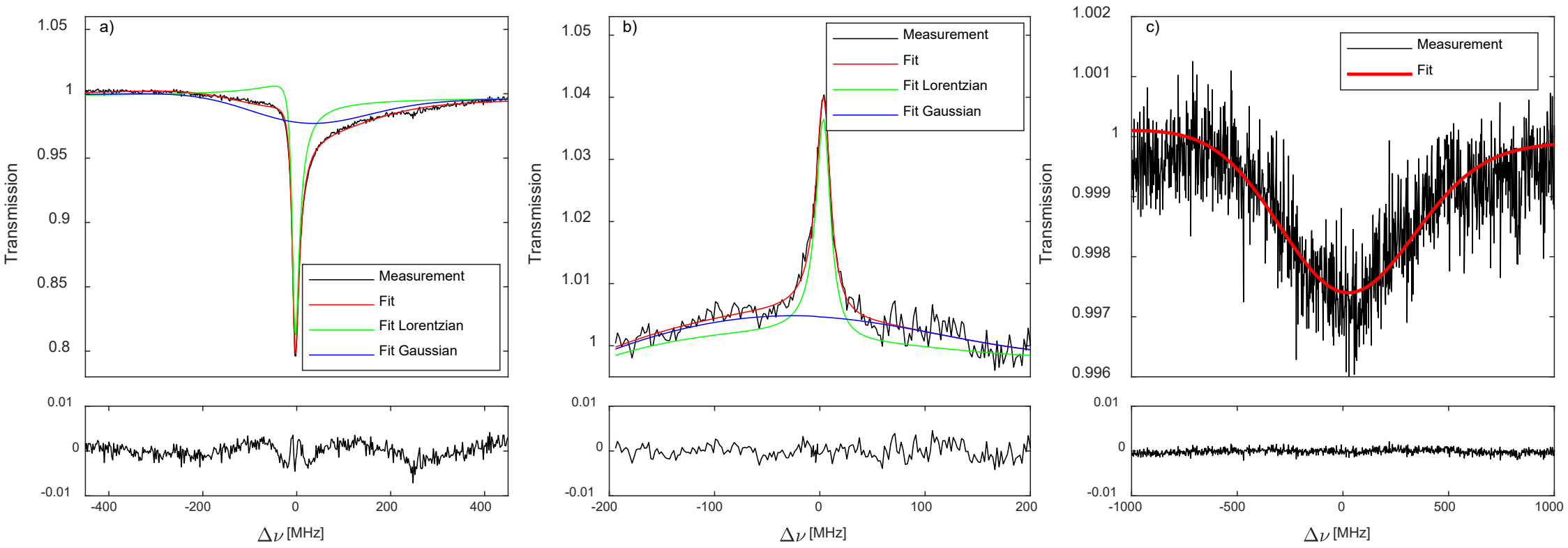

Fig. 2. a) Sub-Doppler ladder-type transition at 6061.335288(5) $cm^{-1}$ (black) and the fit (red curve) plotted as a function of the detuning from the transition center. The green and blue lines represent the fitted Lorentzian and Gaussian components, respectively. b) Sub-Doppler V-type transition at 5994.99534(1) $cm^{-1}$ (black), along with the fit (red curve) and its Lorentzian (green line) and Gaussian (blue line) components. c) Doppler-broadened collision-induced 4LDR line at 5818.877(1) $cm^{-1}$ (black) and the fit curve (red). The bottom panels display the fit residuals.

### 3.2 *Uncertainty in line position and intensity*

The center wavenumber uncertainty arises from the line-fitting process, the optimization of the FTS reference laser wavelength $\lambda_{ref}$ for optical path difference calibration, and the pressure shift of the lines.

The fit uncertainties varied between 56 kHz to 3.7 MHz for the ladder-type transitions, between 0.3 MHz to 1.8 MHz for the V-type transitions, and between 2.6 MHz to 150 MHz for the 4LDR transitions. The fit uncertainty of the Doppler-broadened 4LDR lines is higher than that of the sub-Doppler OODR lines due to their lower SNR.

The sub-nominal sampling-interleaving method relies on matching the FTS sampling points to the comb mode frequencies by minimizing the instrumental line shape (ILS) effects, where the ILS originates from optical path difference (OPD) truncation in the FTS [38, 39]. The value of $\lambda_{ref}$ was determined in post processing by adjusting it to minimize the ILS of absorption lines [39]. Since the sub-Doppler line is always located at the top of the collision-induced Doppler-broadened component whose width is comparable to the $f_{rep}$ of the comb, the center frequency of the sub-Doppler line will be affected by the ILS of the Doppler-broadened component [39], and thus depends on the accuracy of $\lambda_{ref}$. We found the optimum $\lambda_{ref}$, as the mean of the optima of the four strongest ladder-type lines in each measurement while the standard deviation was taken to be the uncertainty. The dependence of the center frequency shift on $\lambda_{ref}$ was evaluated by fitting the ladder-type lines across a range of $\lambda_{ref}$ values. The resulting uncertainty from the sub-nominal analysis was calculated by multiplying the maximum dependence of the center frequency on $\lambda_{ref}$ by the uncertainty of $\lambda_{ref}$. In this way, the center wavenumber uncertainty stemming from sub-nominal analysis was found to be 370 kHz in measurement #1 and 66 kHz in the measurement #2. The larger uncertainty in measurement #1 primarily results from the greater uncertainty in the optimal HeNe wavelength, which in turn is due to the relatively weak ladder-type transitions used to evaluate the optimal HeNe wavelength in that measurement.

Based on the pressure shift coefficient of the $CH_4$ lines in the 6000 $cm^{-1}$ range [44], the pressure shift of the probe lines was calculated to be 136 kHz at 200 mTorr, although we believe this is an overestimation as discussed in Ref. [34].

The uncertainty of integrated absorption for the OODR transitions was a combination of the fit uncertainty and a 5% uncertainty associated with the retrieved cavity enhancement factor.

# 4. Results

## *4.1 Sub-Doppler ladder-type line parameters and assignment*

Fig. 3 shows the comparison of the integrated absorption coefficients and positions of the measured ladder-type lines (black sticks) to the Einstein A-coefficients and positions predicted from the effective Hamiltonian (Fig. 3a), red sticks) and ExoMol (Fig. 3b), blue sticks), respectively. The predicted A-coefficients are plotted inverted for clarity. The reason for some lines being absent from the measurement but appearing in the predictions is attributed to overlap with Doppler-broadened transitions or to their intensities being too weak for detection.

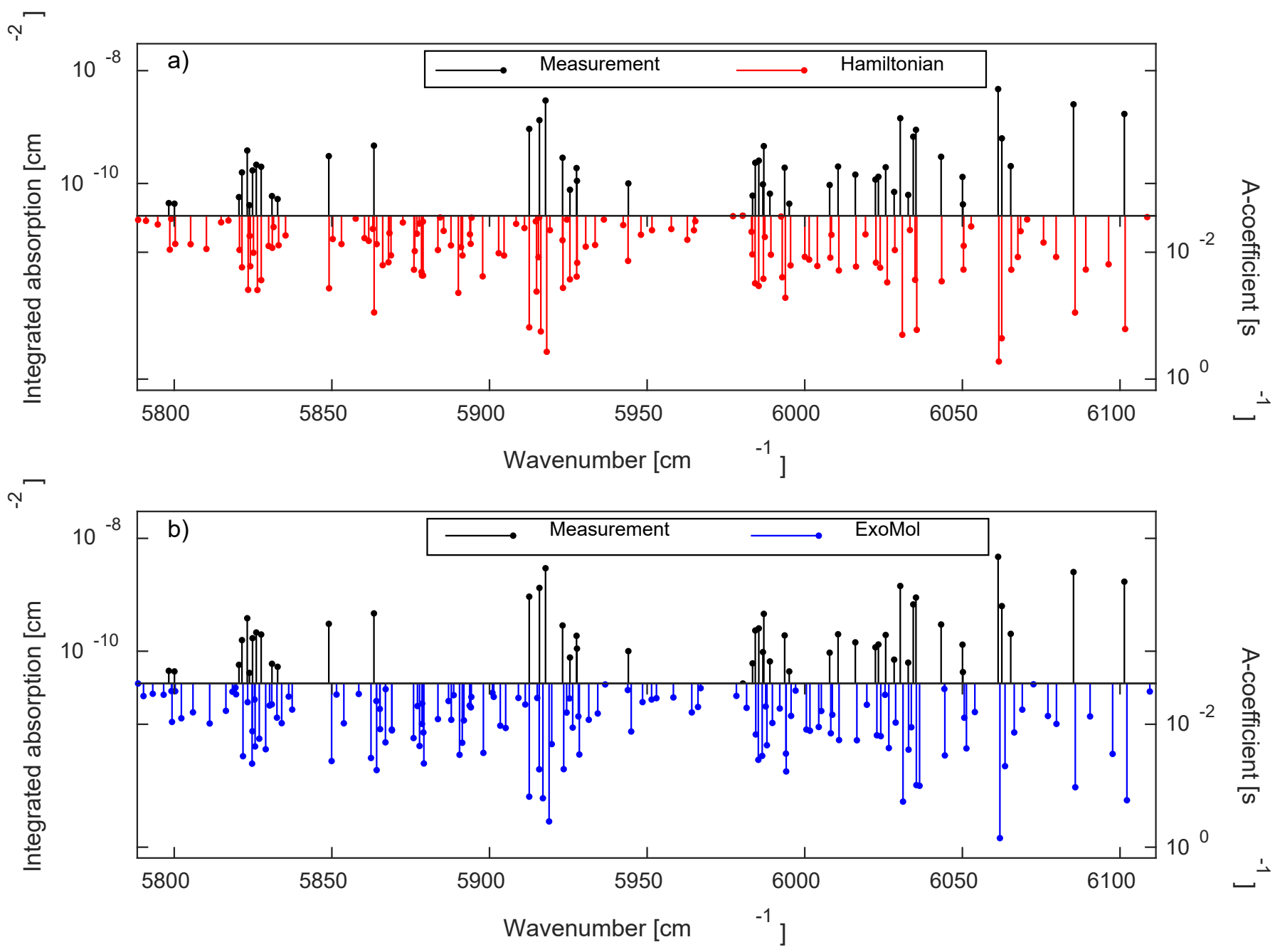

Fig. 3. Integrated absorption coefficients of the observed ladder-type lines (black) compared to the Einstein A-coefficients from a) the effective Hamiltonian (red) and b) ExoMol (blue). The Einstein-A coefficients were plotted inverted for clarity.

We have assigned all observed ladder-type transitions to both the effective Hamiltonian and ExoMol by comparing the wavenumber differences and relative intensities between the experiments and the predictions. It was straightforward to assign the transitions to the Hamiltonian due to its good agreement with the measurements. For some ambiguous assignments in ExoMol, the upper rotational quantum numbers of the transitions assigned to the Hamiltonian were used as a reference to select the most plausible match. Fig. 4a) shows the wavenumber differences between the observed lines and the corresponding assigned transitions from the Hamiltonian and ExoMol. The wavenumber uncertainty is negligible compared to the scatter. Table 1 shows the mean wavenumber discrepancies and the corresponding standard deviations. As observed for the previous $J'' = (7, A_2^{(1)})$ pumped measurements [24], there is an overall negative offset in the predicted wavenumbers, which is more significant for ExoMol and is 40% larger than the ladder-type transitions of the $J'' = (7, A_2^{(1)})$ pumped measurements, where $A_2$ is the rovibrational symmetry. The scatter of the wavenumber discrepancies remains comparable to Ref. [24] for both databases.

The normalized experimental line intensities of ladder-type lines were calculated as the ratio of the integrated absorption of the Lorentzian components of the ladder-type transitions to a selected V-type transition in each measurement. We selected the strongest V-type transitions belonging to the same spectral branch as the pump transition at 5868.14329(2) cm⁻¹ and 5911.61198 (2) cm⁻¹ for measurement #1 and #2 respectively, with upper state assignments $(P,J,C,N) = (4,11,A_2,119)$ and $(4,11,A_2,124)$. These experimentally normalized intensities were then compared with predictions based on the Hamiltonian and ExoMol databases, respectively. The predicted normalized intensities were calculated as $I_{N\,calc} = (A_{LT}g'_{LT}/\upsilon_{LT}^2)/(A_{VT}g'_{VT}/\upsilon_{VT}^2)$, where $A_\kappa$ are the Einstein A-coefficients, $g'_\kappa$ are the upper state degeneracies, and $\nu_\kappa$ are the center wavenumbers of the ladder-type $(\kappa = LT)$ and the V-type $(\kappa = VT)$ transitions. The Einstein A-coefficients for ladder-type transitions were obtained from the Hamiltonian and ExoMol, respectively, while the Einstein A-coefficient for the selected V-type transition was taken from the Hamiltonian. The upper state degeneracies can be expressed as $g'_\kappa = (2J+1)s$, where $J$ is the upper rotational number of the ladder-type and the selected V-type transitions, and $s$ is the spin degeneracy factor which equals 5 for $A_1$ and $A_2$ symmetry states. Fig. 4b) shows the normalized intensity ratios between the measurements and the predictions calculated based on the Hamiltonian and ExoMol. The error bars include the 1σ uncertainties of the Lorentzian components of both the ladder-type and V-type transitions, as well as the uncertainties of the cavity finesse. The mean normalized intensity ratios and the corresponding standard deviations are shown in Table 1. The mean normalized intensity ratios deviate more significantly from unity than those for the $J'' = (7, A_2^{(1)})$ pumped measurements, with the corresponding standard deviations being approximately twice as large. This seems to indicate a certain reduction in prediction accuracy for transitions at higher $J$ numbers.

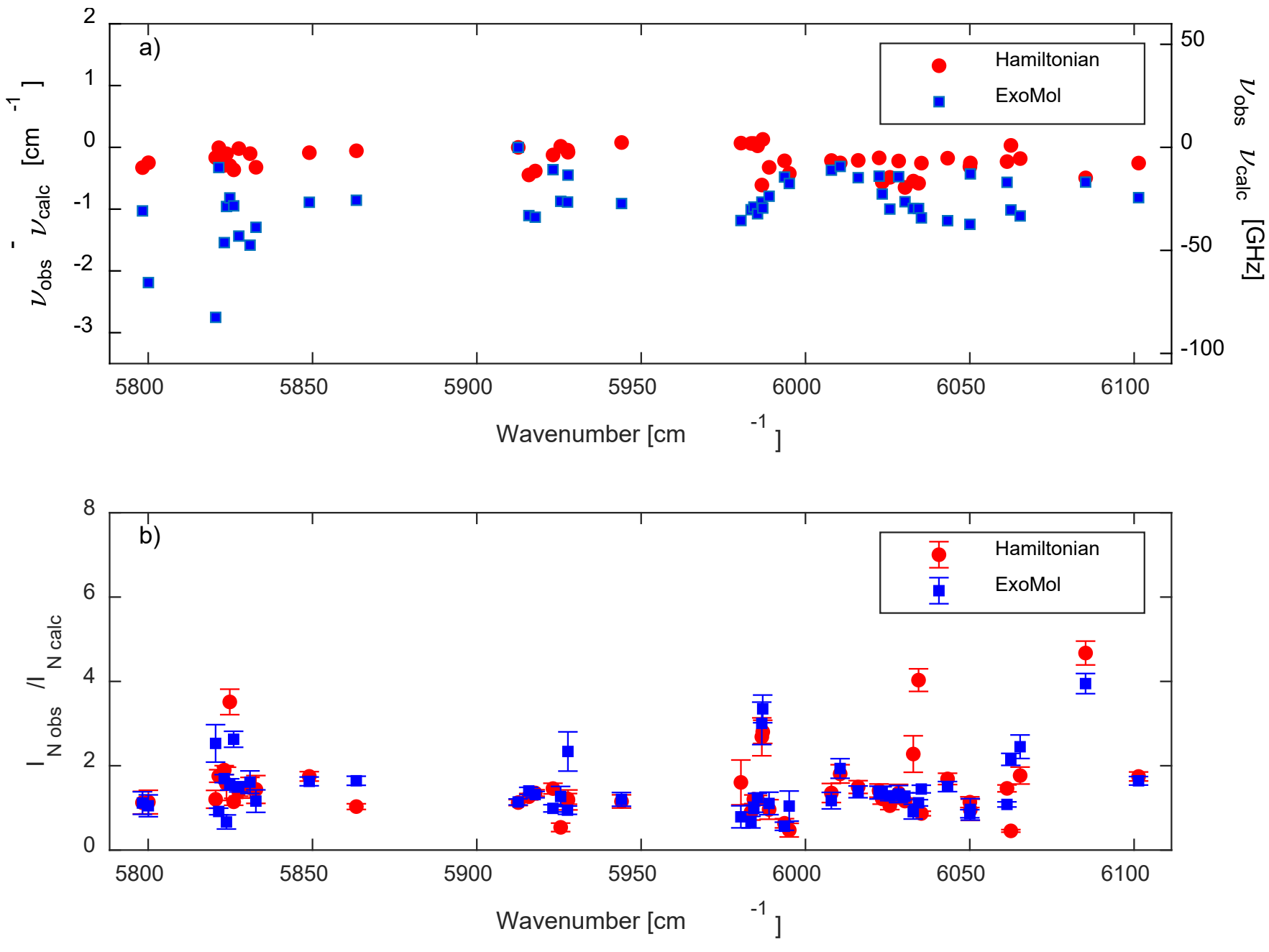

Fig. 4. (a) Wavenumber differences between the observed ladder-type transitions and the effective Hamiltonian (red dots) and ExoMol (blue squares). (b) The normalized intensity ratios between measurements and calculations from Hamiltonian and ExoMol. The error bar represents 1σ uncertainty, which is negligible in Fig.4 (a).

Table 1. Mean values and standard deviations of discrepancies in ladder-type center wavenumbers and normalized intensity ratios between the observations and the predictions from Hamiltonian and ExoMol.

| Reference source | Center wavenumbers [$cm^{-1}$] | | Normalized intensities | |
|---|---|---|---|---|
| | Mean offset | Standard deviation | Mean ratio | Standard deviation |
| Hamiltonian | -0.22 | 0.20 | 1.51 | 0.82 |
| ExoMol | -0.92 | 0.47 | 1.48 | 0.69 |

Table 4 in the Appendixes lists the parameters of the observed ladder-type transitions and their corresponding assignments in the Hamiltonian. Although the Hamiltonian predictions generally agree well with the experiment results, three assignments remain ambiguous since these ladder-type lines have multiple possible matches in the Hamiltonian predictions. We select the most likely assignment options for these three ladder-type transitions by jointly considering the wavenumber differences and intensity ratios between experiments and predictions. These lines are marked with an asterisk on the Hamiltonian transition wavenumbers in the fifth column of the Table 4. The experimental upper state term values were calculated as the sum of the experimental ladder-type wavenumbers, the pump wavenumber 2894.99609799(7) reported by Abe *et al*. [13], and the ground state term value 815.1437445(5) $cm^{-1}$ from our recent work [45]. The final uncertainty was obtained by combining contributions from these three sources, with the dominant contribution arising from the uncertainty in the experimental ladder-type wavenumbers.

The 7 ladder-type transitions reaching 6 vibrational states that have not been experimentally verified in ExoMol are listed in Table 2, where the vibrational quantum numbers were obtained from ExoMol [23]. The mean wavenumber differences between the experiments and ExoMol predictions are largely consistent for states both with and without MARVELized levels.

Table 2. Observed transitions to vibrational states without MARVELized levels. Column 1 lists the experimental center wavenumbers. Column 2 lists the experimental upper state term values. Column 3 provides the corresponding predicted wavenumbers. The quantum numbers are taken from ExoMol [23], where $n_1$, $n_2$, $n_3$ and $n_4$ denote the normal mode vibrational quantum numbers, $l_2$, $l_3$ and $l_4$ represent the vibrational angular momenta quantum numbers, $m_3$ and $m_4$ are the multiplicity index quantum numbers, and $C_v$ indicates the symmetry of the upper vibrational state. The last column indicates the upper state assignments.

| Experimental transition wavenumber [$cm^{-1}$] | Experimental upper state term value [$cm^{-1}$] | ExoMol transition wavenumber [$cm^{-1}$] | $n_1$ | $n_2$ | $l_2$ | $n_3$ | $l_3$ | $m_3$ | $n_4$ | $l_4$ | $m_4$ | $C_v$ | Upper state assignment (*P,J,C,N*) |
|---|---|---|---|---|---|---|---|---|---|---|---|---|---|
| 5821.48296(2) | 9531.62280(2) | 5821.808992 | 1 | 0 | 0 | 2 | 0 | 0 | 0 | 0 | 0 | $A_1$ | (6,12,A1,490) |
| 5823.15709(1) | 9533.29693(1) | 5824.698884 | | | | | | | | | | | (6,12,A1,493) |
| 5986.73146(4) | 9696.87130(4) | 5987.617535 | 1 | 2 | 2 | 1 | 1 | 1 | 0 | 0 | 0 | $F_2$ | (6,12,A1,592) |
| 5987.0150(1) | 9697.1549(1) | 5987.997857 | 1 | 2 | 0 | 1 | 1 | 1 | 0 | 0 | 0 | $F_2$ | (6,12,A1,593) |
| 6025.67217(1) | 9735.81201(1) | 6026.670856 | 1 | 2 | 2 | 1 | 1 | 1 | 0 | 0 | 0 | $F_1$ | (6,12,A1,616) |
| 6050.03540(3) | 9760.17524(3) | 6051.280833 | 0 | 3 | 3 | 1 | 1 | 1 | 1 | 1 | 1 | E | (6,12,A1,626) |
| 6062.514483(8) | 9772.654326(8) | 6063.527021 | 0 | 5 | 1 | 0 | 0 | 0 | 1 | 1 | 1 | $F_1$ | (6,12,A1,630) |

## *4.2 Sub-Doppler V-type line parameters and assignment*

We assigned the 19 observed sub-Doppler V-type transitions to the effective Hamiltonian and ExoMol line lists. The assignments were straightforward as the lower states of the transitions are known. Fig. 5 shows the wavenumber differences between the observed V-type lines and the corresponding predictions from the Hamiltonian and ExoMol. In addition, we assigned the V-type transitions to lines in the WKLMC (Wang, Kassi, Leshchishina, Mondelain, Campargue) [46] and HITRAN2024 [47] databases, but only 2 assignments were unambiguous because they had assigned lower states in both databases. For the remaining assignments in the WKLMC and HITRAN2024, we selected only those lines whose wavenumbers and intensities were closest to those of the observed V-type transitions. Since WKLMC does not cover the entire probe range in our measurements, only 14 observed V-type transitions could be assigned to this database. Table 3 lists the mean values and standard deviations of the wavenumber discrepancies between the experiment and Hamiltonian, ExoMol, WKLMC, and HITRAN. For the Hamiltonian, the mean wavenumber offset and standard deviation are 5.7 and 4 times larger, respectively, than those of the previously observed V-type transitions from $J'' = (7, A_2^{(1)})$ pumped measurements [24]. For ExoMol, the increase in the discrepancies is even larger with the mean offset and standard deviation larger compared to Ref. [24] by factors of 26 and 57, respectively. One major reason for this is that, while in Ref. [24] all the observed V-type lines are MARVELized, in the present work 9 of the observed V-type transitions are purely *ab initio*. For the remaining 10 MARVELized V-type transitions, the mean offset and standard deviation are reduced to 6 and 10 times those reported in Ref. [24] For the WKLMC and HITRAN, the mean wavenumber offset and standard deviation are comparable to the V-type transitions in $J'' = (7, A_2^{(1)})$ pumped measurements [24]. Table 5 in the Appendixes shows the parameters of the observed V-type transitions and the corresponding assignments to the Hamiltonian. The transitions marked with an asterisk in the first column of Table 5 are those used to calculate the normalized ladder-type transition intensities.

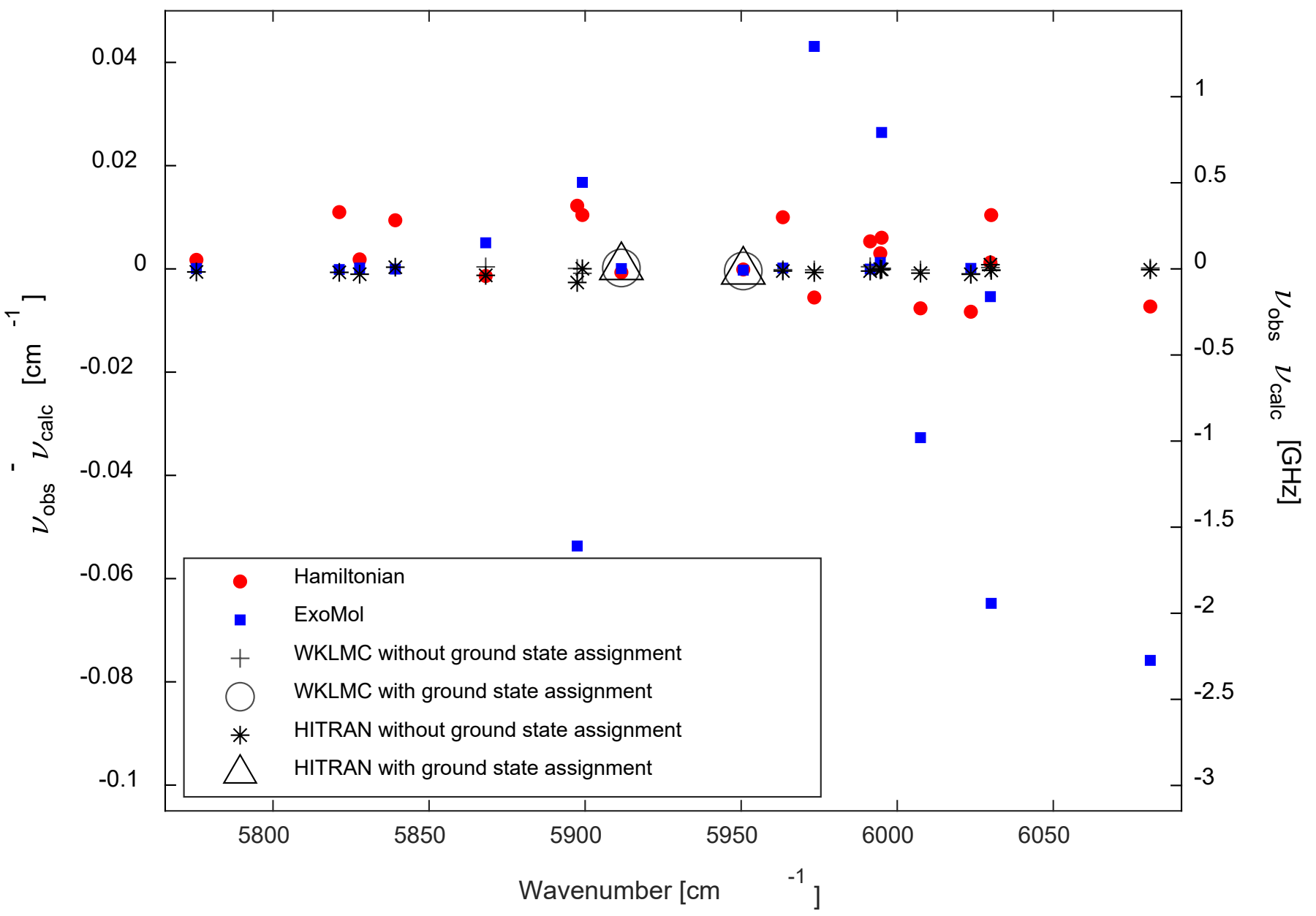

Fig. 5. Comparison of the observed V-type transition center wavenumbers to the effective Hamiltonian (red circles), ExoMol (blue squares), WKLMC (grey plus signs and circles), and HITRAN (black asterisks and triangles) line lists. The experimental uncertainties are negligible compared to the scatter.

Table 3. Mean values and standard deviations of the V-type center wavenumber differences between the experiment and Hamiltonian, ExoMol, WKLMC and HITRAN.

| Reference | Mean offset [cm$^{-1}$] | Standard deviation [cm$^{-1}$] |
|---|---|---|
| Hamiltonian | $2.7 \cdot 10^{-3}$ | $6.8 \cdot 10^{-3}$ |
| ExoMol | $-7.4 \cdot 10^{-3}$ | $3.0 \cdot 10^{-2}$ |
| WKLMC | $-1.1 \cdot 10^{-4}$ | $4.1 \cdot 10^{-4}$ |
| HITRAN | $-4.8 \cdot 10^{-4}$ | $7.4 \cdot 10^{-4}$ |

### 4.3 *Doppler-broadened 4-level line parameters*

Fig. 6 shows the integrated absorption coefficients of the collision-induced 4LDR lines in the current P(12, $A_1^{(2)}$) (black sticks) and previous $J''$=(7, $A_2^{(1)}$) [35] pumped measurements (red sticks). The integrated absorption from the previous measurement is plotted inverted for clarity. The integrated absorption in the current measurement is lower than that in the previous measurement, which is mainly due to the different intensities of the pump transitions. The line intensities of the $J''$=(7, $A_2^{(1)}$) pump lines are in the range of $8.41\times10^{-20}$ to $1.35\times10^{-19}$ cm$^{-1}$/(molec·cm$^{-2}$), while the line intensity of the current P(12, $A_1^{(2)}$) pump line is $1.11\times10^{-20}$ cm$^{-1}$/(molec·cm$^{-2}$) [47]. Out of the 170 4LDR lines observed here, we identified 163 lines in common with previously [35] as those with a wavenumber difference less than 0.0076 cm$^{-1}$. This means that most of the pumped population was transferred to the same states as when pumping levels with $J''$=7, with only 7 new lines appearing when pumping the $J''$=12 level. The standard deviation of the wavenumber differences for spectral lines in common between the two datasets was used to reflect the combined uncertainty of the measurement sets. We then calculate the uncertainty of the new 4LDR lines to be 0.0015 $cm^{-1}$, based on the calculated combined uncertainty together with the 0.0013 $cm^{-1}$ uncertainty of the 4LDR lines in the $J''$=(7, $A_2^{(1)}$) [35] pumped measurements, the latter being obtained from the agreement in combination differences. We could not identify combination differences involving the new 4LDR lines and thus did not make any new assignments. Table 6 in the Appendixes displays the parameters of the newly observed 4LDR lines.

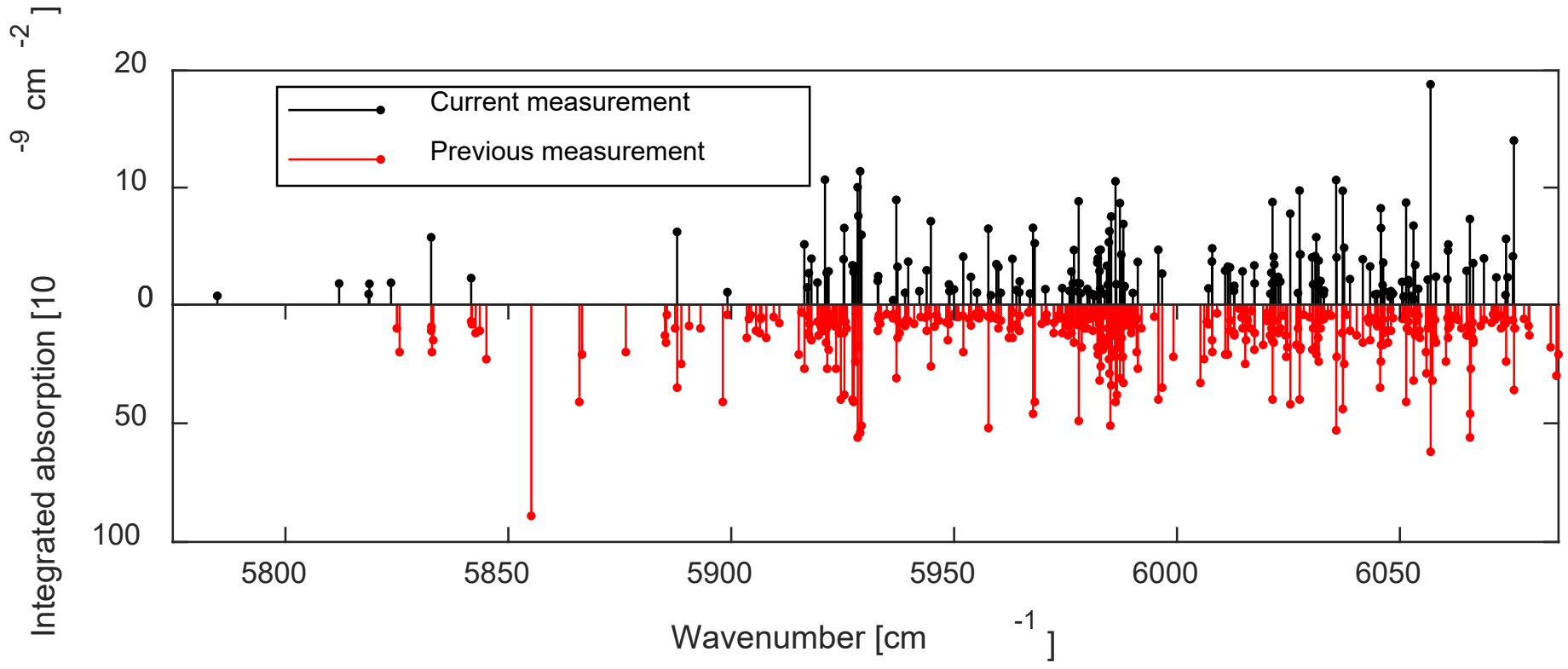

Fig. 6. Integrated absorption coefficient of the collision-induced 4LDR transitions when pumping P(12, $A_1^{(2)}$) (black) and $J''$ = (7, $A_2^{(1)}$) (red) states.

## 5 Conclusion

We used a cavity-enhanced comb-based OODR spectrometer to measure and assign sub-Doppler ladder-type transitions to states in the *P*6 polyad around 9510-9810 $cm^{-1}$ and V-type transitions to *P*4 levels around 6590-6896 $cm^{-1}$ with rotational quantum number $J \geq 10$. All 49 ladder-type transitions and 19 V-type transitions were assigned by comparison with theoretical predictions from an effective Hamiltonian [8] and the ExoMol [23] database. Via the ladder-type transitions, we probed 6 vibrational states in the *P*6 polyad that have not yet been included in the MARVEL procedure [23]. We also assigned the observed V-type transitions to the WKLMC and HITRAN, however, only 2 assignments are unambiguous as the others lack assigned lower states of the rotational levels $J$ = 11-13 in both databases. For both ladder-type and V-type transitions, we observed larger discrepancies with both the Hamiltonian and ExoMol compared to our previous work pumping the $J''$=(7, $A_2^{(1)}$) ground state. In particular, the mean wavenumber difference and standard deviation between the currently observed V-type lines and the ExoMol predictions are 26 and 57 times larger, respectively, than the previous $J''$=(7, $A_2^{(1)}$) pumped measurement. This is mainly due to that nearly half of the observed V-type transitions are unMARVELized lines. In addition, the scatter of the normalized intensities calculated based on the effective Hamiltonian is slightly greater than that calculated based on ExoMol, and the mean normalized intensity calculated based on both databases is 1.5. We also report 7 new Doppler-broadened 4LDR transitions induced by the redistribution of the pumped population due to collisional transfer.

## CRediT authorship contribution statement

**Yuan Cao:** Writing – original draft, Visualization, Methodology, Investigation, Formal analysis, Data curation. **Adrian Hjältén:** Writing – original draft, Investigation, Data curation. **Vinicius Silva de Oliveira:** Methodology, Data curation. **Isak Silander:** Formal analysis, Data curation. **Michael Rey:** Validation, Formal analysis. **Kevin K. Lehmann:** Writing – review & editing, Validation,

Conceptualization. **Aleksandra Foltynowicz:** Writing – review & editing, Supervision, Resources, Project administration, Funding acquisition, Conceptualization.

## Declaration of competing interest

The authors declare that they have no known competing financial interests or personal relationships that could have appeared to influence the work reported in this paper.

## Acknowledgments

This project is supported by the Knut and Alice Wallenberg Foundation (grant: KAW 2020.0303), and the Swedish Research Council (grant: 2020-00238). Y. C. acknowledges funding from the Wenner Gren Foundation (grant: UPD2024-0201). K.K.L. acknowledges funding from the U.S. National Science Foundation (grant: CHE-2108458) and the Wenner Gren Foundation (grant: GFOv2024-0010). The authors thank Piotr Masłowski for the loan of the rubidium clock and Sergey Yurchenko for providing a list of unMARVELized states.

## Data availability

Data will be made available on request.

# Appendixes

The Appendixes contain three tables of the parameters and assignments of the observed OODR transitions. The first table lists the line positions, upper state term values, integrated absorption coefficients, and Lorentzian line widths of the experimental ladder-type transitions compared to the predicted wavenumbers, Einstein A-coefficients, upper vibrational states and upper state quantum numbers (*P,J,C,N*) from the effective Hamiltonian. The second table lists the line positions, integrated absorption coefficients, and Lorentzian line widths of the experimental V-type transitions, together with their corresponding assignment results from the Hamiltonian predictions, including the predicted wavenumbers, line intensities, upper vibrational states, and upper state quantum numbers (*P,J,C,N*). The third table lists the experimental line positions and integrated absorption coefficients of the 7 newly observed Doppler-broadened 4LDR transitions.

Table 4. Parameters of the ladder-type lines in $\nu_3$ P(12, $A_1^{(2)}$) pumped measurement, i.e., starting from the $J'$ = (11, $A_2^{(15)}$) level of the *P*2 polyad with a term value 2894.99609799(7) + 815.1437445(5) = 3710.1398425(5) cm$^{-1}$. Column 1 lists the experimental center wavenumber and the corresponding uncertainty. Column 2 lists the final state term value derived from the sum of the probe transition and the $J'$ = (11, $A_2^{(15)}$) term value. Column 3 lists the integrated absorption from the fitted Lorentzian component. Column 4 shows the half width at half maximum (HWHM) of the fitted Lorentzian component. Column 5-8 show the Hamiltonian model predicted transition wavenumber, Einstein A-coefficient, upper state vibrational assignment, and upper state quantum numbers (*P,J,C,N*), where *P* is polyad, *J* total angular momentum, *C* the rovibrational symmetry, and *N* the counting number distinguishing states within a (*P,J,C*)-block in ascending order. The predicted spectral lines marked with an asterisk in the fifth column represent the ambiguous assignment results.

| 1 | 2 | 3 | 4 | 5 | 6 | 7 | 8 |
|---|---|---|---|---|---|---|---|
| Experimental transition wavenumber [cm$^{-1}$] | Experimental upper state term value [cm$^{-1}$] | Integrated absorption [$10^{-9}$ cm$^{-2}$] | Lorentzian HWHM [MHz] | Hamiltonian transition wavenumber [cm$^{-1}$] | Hamiltonian Einstein A-coefficient [s$^{-1}$] | Upper vibrational state | Upper state assignment (*P,J,C,N*) |
| 5798.26820(5) | 9508.40804(5) | 0.045(7) | 7(1) | 5798.596048 | 0.009105 | 0 0 3 0 1F1 | (6,10,A1,527) |
| 5800.06757(4) | 9510.20741(4) | 0.044(7) | 5(1) | 5800.316617 | 0.00738 | 0 0 2 2 5F2 | (6,12,A1,476) |
| 5820.53379(5) | 9530.67363(5) | 0.057(6) | 10(2) | 5820.699546 | 0.009196 | 1 1 1 1 1F1 | (6,12,A1,491) |
| 5821.48296(2) | 9531.62280(2) | 0.157(9) | 8.6(5) | 5821.488718 | 0.01728 | 1 3 0 1 1F2 | (6,12,A1,492) |
| 5823.15709(1) | 9533.29693(1) | 0.38(2) | 7.8(2) | 5823.449192 | 0.03911 | 1 1 1 1 1A1 | (6,12,A1,493) |
| 5823.83826(5) | 9533.97810(5) | 0.041(6) | 7(1) | 5823.941988* | 0.005518 | 0 1 2 1 1F1 | (6,11,A1,504) |
| 5824.83464(3) | 9534.97448(3) | 0.17(1) | 10.6(7) | 5825.130073 | 0.01018 | 1 0 2 0 1E | (6,11,A1,505) |
| 5826.02317(2) | 9536.16301(2) | 0.21(1) | 7.6(3) | 5826.387291 | 0.0394 | 1 0 2 0 1E | (6,11,A1,506) |
| 5827.57293(2) | 9537.71277(2) | 0.20(1) | 7.1(4) | 5827.592847 | 0.02754 | 0 1 2 1 2E | (6,12,A1,496) |
| 5831.00075(4) | 9541.14059(4) | 0.060(6) | 9(1) | 5831.102332 | 0.008521 | 0 1 2 1 1F1 | (6,11,A1,508) |
| 5832.79256(4) | 9542.93241(4) | 0.053(9) | 6(1) | 5833.115122 | 0.007763 | 1 0 2 0 1E | (6,11,A1,510) |
| 5849.02802(2) | 9559.16787(2) | 0.31(2) | 8.2(4) | 5849.114083 | 0.03703 | 0 1 2 1 5F1 | (6,11,A1,518) |
| 5863.34625(3) | 9573.48609(3) | 0.47(5) | 8.0(7) | 5863.401776 | 0.08894 | 0 1 2 1 4F1 | (6,12,A1,518) |
| 5912.60096(1) | 9622.74080(1) | 0.93(5) | 7.8(2) | 5912.60134 | 0.1528 | 0 0 3 0 2F2 | (6,10,A1,560) |
| 5915.837321(9) | 9625.977164(9) | 1.32(7) | 8.0(2) | 5916.284008 | 0.1775 | 0 0 3 0 1F1 | (6,11,A1,557) |
| 5917.772183(6) | 9627.912026(6) | 3.0(1) | 7.9(1) | 5918.155826 | 0.3709 | 0 0 3 0 1F1 | (6,11,A1,558) |
| 5923.18075(2) | 9633.32059(2) | 0.29(2) | 8.7(7) | 5923.303911 | 0.03652 | 0 2 2 0 1E | (6,10,A1,563) |
| 5925.52164(3) | 9635.66148(3) | 0.08(1) | 5(1) | 5925.504786 | 0.02674 | 0 2 2 0 1E | (6,10,A1,564) |
| 5927.60905(3) | 9637.74890(3) | 0.19(2) | 7.0(7) | 5927.654836 | 0.02422 | 0 1 2 1 4F1 | (6,12,A1,558) |
| 5927.74139(6) | 9637.88123(6) | 0.11(3) | 7(2) | 5927.820075 | 0.01466 | 0 1 2 1 5F2 | (6,12,A1,559) |
| 5944.04029(3) | 9654.18013(3) | 0.100(9) | 7.4(9) | 5943.961845 | 0.01369 | 0 1 2 1 5F1 | (6,12,A1,566) |
| 5980.36423(3) | 9690.50408(3) | 0.027(5) | 4(1) | 5980.295456 | 0.002662 | 0 5 0 1 3F2 | (6,12,A1,587) |
| 5983.44907(3) | 9693.58892(3) | 0.061(9) | 5.3(9) | 5983.385444 | 0.01081 | 0 1 2 1 5F1 | (6,12,A1,590) |

| 5984.30670(1) | 9694.44655(1) | 0.23(1) | 7.5(4) | 5984.243847 | 0.03092 | 0 1 2 1 4F2 | (6,12,A1,591) |
|---|---|---|---|---|---|---|---|
| 5985.42302(2) | 9695.56286(2) | 0.25(2) | 9.1(5) | 5985.397892 | 0.03397 | 0 1 2 1 5F1 | (6,12,A1,592) |
| 5986.73146(4) | 9696.87130(4) | 0.10(1) | 6(1) | 5987.34153 | 0.005746 | 0 1 2 1 4F2 | (6,12,A1,594) |
| 5987.0150(1) | 9697.1549(1) | 0.46(8) | 24(4) | 5986.886016 | 0.0262 | 0 0 3 0 2F2 | (6,12,A1,593) |
| 5988.94042(5) | 9699.08026(5) | 0.07(1) | 5(1) | 5989.263736 | 0.01094 | 0 3 1 1 4F1 | (6,12,A1,595) |
| 5993.62628(4) | 9703.76612(4) | 0.19(6) | 4(1) | 5993.844352 | 0.05216 | 0 0 3 0 2F2 | (6,11,A1,575) |
| 5995.07521(3) | 9705.21505(3) | 0.04(1) | 2.0(7) | 5995.497425 | 0.01613 | 0 2 2 0 1F2 | (6,11,A1,576) |
| 6007.89076(5) | 9718.03060(5) | 0.09(1) | 9(2) | 6008.103851* | 0.01216 | 0 2 2 0 1F1 | (6,11,A1,580) |
| 6010.56672(5) | 9720.70657(5) | 0.20(3) | 10(1) | 6010.818313 | 0.01936 | 0 2 2 0 2F2 | (6,11,A1,581) |
| 6016.04186(2) | 9726.18170(2) | 0.143(9) | 8.7(5) | 6016.252869 | 0.01684 | 0 2 2 0 1F2 | (6,11,A1,583) |
| 6022.41037(2) | 9732.55022(2) | 0.117(8) | 8.4(6) | 6022.580725 | 0.01462 | 1 4 0 0 1E | (6,11,A1,586) |
| 6023.35962(2) | 9733.49947(2) | 0.131(8) | 8.2(4) | 6023.922502 | 0.01751 | 1 2 1 0 2F2 | (6,12,A1,615) |
| 6025.67217(1) | 9735.81201(1) | 0.19(1) | 7.4(3) | 6026.156972 | 0.02984 | 0 3 1 1 3F1 | (6,12,A1,617) |
| 6028.35390(3) | 9738.49374(3) | 0.071(6) | 7.9(9) | 6028.574612 | 0.009239 | 1 4 0 0 1E | (6,11,A1,588) |
| 6030.290058(6) | 9740.429901(6) | 1.43(7) | 7.85(7) | 6030.938759 | 0.2 | 0 0 3 0 1F1 | (6,12,A1,618) |
| 6032.81657(4) | 9742.95641(4) | 0.063(8) | 9(1) | 6033.362698* | 0.004472 | 0 0 3 0 1F1 | (6,12,A1,620) |
| 6034.41556(1) | 9744.55540(1) | 0.68(4) | 8.1(3) | 6034.996333 | 0.02723 | 0 3 1 1 4F1 | (6,12,A1,621) |
| 6035.275149(6) | 9745.414991(6) | 0.89(5) | 8.3(1) | 6035.530013 | 0.1674 | 0 0 3 0 1F1 | (6,12,A1,622) |
| 6043.25314(2) | 9753.39298(2) | 0.30(2) | 10.6(6) | 6043.428397 | 0.02872 | 0 3 1 1 1F1 | (6,12,A1,624) |
| 6050.03540(3) | 9760.17524(3) | 0.13(1) | 8.7(8) | 6050.348979 | 0.01878 | 0 3 1 1 4F2 | (6,12,A1,627) |
| 6050.16804(3) | 9760.30789(3) | 0.043(7) | 4(1) | 6050.419949 | 0.007921 | 0 2 2 0 1A2 | (6,11,A1,595) |
| 6061.335288(5) | 9771.475131(5) | 4.7(2) | 8.17(5) | 6061.565526 | 0.5279 | 0 0 3 0 1A1 | (6,12,A1,630) |
| 6062.514483(8) | 9772.654326(8) | 0.63(3) | 7.4(2) | 6062.482538 | 0.2269 | 0 5 0 1 1F1 | (6,12,A1,631) |
| 6065.33509(4) | 9775.47494(4) | 0.20(2) | 10(1) | 6065.516612 | 0.01886 | 0 3 1 1 4F1 | (6,12,A1,632) |
| 6085.24747(7) | 9795.38731(7) | 2.5(3) | 20(3) | 6085.741372 | 0.08945 | 0 2 2 0 2E | (6,12,A1,640) |
| 6101.37451(8) | 9811.51436(8) | 1.7(1) | 24(3) | 6101.628765 | 0.1624 | 0 2 2 0 1A1 | (6,12,A1,643) |

Table 5. Parameters of the observed V-type transitions and their comparison with Hamiltonian. Column 1 lists the measured center wavenumber. Column 2 lists the integrated absorption from the fitted Lorentzian component. Column 3 shows the HWHM of the fitted Lorentzian component. Column 4, 5, 6, and 7 list the predicted wavenumber, line intensity, upper state vibrational assignment, and upper state quantum numbers (*P,J,C,N*) in the Hamiltonian. In both measurements, the strongest V-type experimental transitions marked with an asterisk, belonging to the same spectral branch as the pump transition, were used to calculate the corresponding normalized intensity. The experimental transitions marked with "Ca" represent the unMARVELized lines.

| 1 | 2 | 3 | 4 | 5 | 6 | 7 |
|---|---|---|---|---|---|---|
| Experimental transition wavenumber [$cm^{-1}$] | Integrated absorption [$10^{-9}$ $cm^{-2}$] | Lorentzian HWHM [MHz] | Hamiltonian transition wavenumber [$cm^{-1}$] | Hamiltonian line intensity [$cm^{-1}$/(mol $cm^{-2}$)] | Upper vibrational state | Upper state Assignment (*P,J,C,N*) |
| 5775.45400(4) | 0.14(3) | 7(1) | 5775.452253 | $3.378\cdot10^{-24}$ | 0 0 1 2 3F2 | (4,13,A2,72) |
| 5821.22973(4) | 0.08(1) | 7(1) | 5821.21873 | $1.728\cdot10^{-24}$ | 0 0 1 2 2F1 | (4,13,A2,79) |
| 5827.75284(5) | 0.10(2) | 9(1) | 5827.750996 | $2.063\cdot10^{-24}$ | 0 0 1 2 3F2 | (4,13,A2,81) |
| 5839.17539(3) | 0.16(2) | 6.4(8) | 5839.165968 | $4.513\cdot10^{-24}$ | 0 1 1 1 1F1 | (4,12,A2,98) |
| 5868.14329(2)*[,Ca] | 1.4(1) | 7.7(6) | 5868.144664 | $4.252\cdot10^{-23}$ | 0 0 2 0 1F2 | (4,11,A2,119) |
| 5897.46119(6) [Ca] | 0.18(3) | 7(2) | 5897.448956 | $4.816\cdot10^{-24}$ | 0 2 0 2 1E | (4,13,A2,94) |
| 5899.11734(3) | 0.20(3) | 5.4(8) | 5899.1069 | $4.997\cdot10^{-24}$ | 0 1 1 1 2F1 | (4,13,A2,95) |
| 5911.61198(2)* | 0.40(3) | 7.9(7) | 5911.612655 | $9.521\cdot10^{-24}$ | 0 0 2 0 1E | (4,11,A2,124) |
| 5950.66510(3) | 0.048(6) | 6.1(7) | 5950.66521 | $1.198\cdot10^{-24}$ | 0 2 1 0 1F1 | (4,11,A2,129) |
| 5963.35094(2) | 0.41(3) | 8.5(5) | 5963.340935 | $8.496\cdot10^{-24}$ | 1 0 1 0 1F2 | (4,13,A2,107) |
| 5973.41830(2) [Ca] | 0.040(4) | 6.1(6) | 5973.423844 | $1.154\cdot10^{-24}$ | 0 1 1 1 2F2 | (4,13,A2,111) |
| 5991.32612(6) | 0.035(6) | 9(2) | 5991.320781 | $8.142\cdot10^{-25}$ | 1 0 1 0 1F2 | (4,13,A2,117) |
| 5994.56667(1) [Ca] | 2.8(2) | 7.2(3) | 5994.563639 | $7.231\cdot10^{-23}$ | 0 0 2 0 1F2 | (4,12,A2,123) |
| 5994.99534(1) [Ca] | 0.75(4) | 8.4(3) | 5994.989305 | $1.385\cdot10^{-23}$ | 0 0 2 0 1F2 | (4,12,A2,124) |
| 6007.44559(3) [Ca] | 0.080(8) | 8.1(7) | 6007.453229 | $2.196\cdot10^{-24}$ | 0 1 1 1 1E | (4,13,A2,119) |
| 6023.58205(3) | 0.069(8) | 8.5(9) | 6023.590352 | $1.331\cdot10^{-24}$ | 0 1 1 1 1F1 | (4,13,A2,121) |
| 6029.84546(1) [Ca] | 0.21(1) | 9.1(4) | 6029.844217 | $4.018\cdot10^{-24}$ | 0 0 2 0 1E | (4,12,A2,127) |
| 6030.10721(1) [Ca] | 0.24(1) | 9.4(3) | 6030.096782 | $5.139\cdot10^{-24}$ | 0 1 1 1 1F2 | (4,13,A2,123) |
| 6081.07585(4) [Ca] | 0.08(2) | 5(1) | 6081.083143 | $2.206\cdot10^{-24}$ | 0 2 1 0 1F2 | (4,12,A2,134) |

Table 6. Parameters of the collision-induced 4LDR transitions. Column 1 and 2 list the measured center wavenumber and integrated absorption coefficient.

| 1 | 2 |
|---|---|
| Experimental wavenumber [$cm^{-1}$] | Experimental integrated absorption [$10^{-9}$ $cm^{-2}$] |
| 5784.7560(15) | 0.75(9) |
| 5812.0737(15) | 1.8(1) |
| 5818.7043(15) | 0.90(8) |
| 5818.8770(15) | 1.8(1) |
| 5823.7120(15) | 1.9(1) |
| 5921.0699(15) | 10.7(7) |
| 5942.2477(15) | 1.15(8) |

## References

[1] Swain M, Vasisht G, Tinetti G. The presence of methane in the atmosphere of an extrasolar planet. Nature. 2008;452:329-31.

[2] Ulenikov ON, Bekhtereva ES, Albert S, Bauerecker S, Niederer HM, Quack M. Survey of the high resolution infrared spectrum of methane ($^{12}CH_4$ and $^{13}CH_4$): Partial vibrational assignment extended towards 12 000 $cm^{-1}$. J Chem Phys. 2014;141.

[3] Wong A, Bernath PF, Rey M, Nikitin AV, Tyuterev VG. Atlas of experimental and theoretical high-temperature methane cross sections from T = 295 to 1000 K in the near-infrared. Astrophys J Suppl Ser. 2019;240:4.

[4] Hargreaves RJ, Gordon IE, Rey M, Nikitin AV, Tyuterev VG, Kochanov RV, et al. An accurate, extensive, and practical line list of methane for the HITEMP database. Astrophys J Suppl Ser. 2020;247:55.

[5] Ghysels M, Vasilchenko S, Mondelain D, Béguier S, Kassi S, Campargue A. Laser absorption spectroscopy of methane at 1000 K near 1.7 μm: A validation test of the spectroscopic databases. J Quant Spectrosc Radiat Transfer. 2018;215:59-70.

[6] Amyay B, Gardez A, Georges R, Biennier L, Vander Auwera J, Richard C, et al. New investigation of the $\nu_3$ C–H stretching region of $^{12}CH_4$ through the analysis of high temperature infrared emission spectra. J Chem Phys. 2018;148.

[7] Nikitin A, Boudon V, Wenger C, Albert S, Brown L, Bauerecker S, et al. High resolution spectroscopy and the first global analysis of the Tetradecad region of methane $^{12}CH_4$. PCCP. 2013;15:10071-93.

[8] Rey M. Novel methodology for systematically constructing global effective models from ab initio - based surfaces: A new insight into high-resolution molecular spectra analysis. J Chem Phys. 2022;156.

[9] Kefala K, Boudon V, Yurchenko SN, Tennyson J. Empirical rovibrational energy levels for methane. J Quant Spectrosc Radiat Transfer. 2024;316:108897.

[10] Ishibashi C, Kourogi M, Imai K, Widiyatmoko B, Onae A, Sasada H. Absolute frequency measurement of the saturated absorption lines of methane in the 1.66 μm region. Opt Commun. 1999;161:223-6.

[11] Okubo S, Nakayama H, Iwakuni K, Inaba H, Sasada H. Absolute frequency list of the $\nu_3$-band transitions of methane at a relative uncertainty level of $10^{-11}$. Opt Express. 2011;19 24:23878-88.

[12] Baumann E, Giorgetta FR, Swann WC, Zolot AM, Coddington I, Newbury NR. Spectroscopy of the methane $\nu_3$ band with an accurate midinfrared coherent dual-comb spectrometer. Phys Rev A. 2011;84:062513.

[13] Abe M, Iwakuni K, Okubo S, Sasada H. Accurate transition frequency list of the $\nu_3$ band of methane from sub-Doppler resolution comb-referenced spectroscopy. J Opt Soc Am B. 2013;30:1027-35.

[14] Kocheril PA, Markus CR, Esposito AM, Schrader AW, Dieter TS, McCall BJ. Extended sub-Doppler resolution spectroscopy of the $\nu_3$ band of methane. J Quant Spectrosc Radiat Transfer. 2018;215:9-12.

[15] Lin H, Yang L, Feng XJ, Zhang JT. Discovery of new lines in the *R*9 multiplet of the $2\nu_3$ band of $^{12}CH_4$. Phys Rev Lett. 2019;122:013002.

[16] Yang L, Lin H, Plimmer M, Feng X-J, Ma Y-J, Luo J-T, et al. Measurement of the spectral line positions in the $2\nu_3$ R(6) manifold of methane. J Quant Spectrosc Radiat Transfer. 2020;245:106888.

[17] Votava O, Kassi S, Campargue A, Romanini D. Comb coherence-transfer and cavity ring-down saturation spectroscopy around 1.65 μm: kHz-accurate frequencies of transitions in the $2\nu_3$ band of $^{12}CH_4$. PCCP. 2022;24:4157-73.

[18] Shah Riyadh S, Telfah H, Jones IW, Bersson JS, Cheng C-F, Hu S-M, et al. Mid-infrared Doppler-free saturation absorption spectroscopy of the Q branch of $CH_4$ $\nu_3 = 1$ band using a rapid-scanning continuous-wave optical parametric oscillator. Opt Lett. 2024;49:4230-3.

[19] Pierre G, Hilico J-C, de Bergh C, Maillard J-P. The region of the $3\nu_3$ band of methane. I Mol Spectrosc. 1980;82:379-93.

[20] Béguier S, Liu A, Campargue A. An empirical line list for methane near 1 μm (9028–10,435 $cm^{-1}$). J Quant Spectrosc Radiat Transfer. 2015;166:6-12.
[21] Nikitin A, Protasevich A, Rey M, Serdyukov V, Sinitsa L, Lugovskoy A, et al. Improved line list of $^{12}CH_4$ in the 8850–9180 $cm^{-1}$ region. J Quant Spectrosc Radiat Transfer. 2019;239:106646.
[22] Campargue A, Karlovets E, Vasilchenko S, Turbet M. The high resolution absorption spectrum of methane in the 10800–14000 $cm^{-1}$ region: literature review, new results and perspectives. PCCP. 2023;25:32778-99.
[23] Yurchenko SN, Owens A, Kefala K, Tennyson J. ExoMol line lists – LVII. High accuracy ro-vibrational line list for methane ($CH_4$). Mon Not R Astron Soc. 2024;528:3719-29.
[24] Hjältén A, Silva de Oliveira V, Silander I, Rosina A, Rey M, Rutkowski L, et al. Measurement and assignment of $J = 5$ to 9 rotational energy levels in the 9070–9370 $cm^{-1}$ range of methane using optical frequency comb double-resonance spectroscopy. J Chem Phys. 2024;161.
[25] Hjältén A, Silva de Oliveira V, Rey M, Silander I, Lehmann KK, Foltynowicz A. Measurement and assignment of E-symmetry states in the 6010-6110 $cm^{-1}$ and 8940-9150 $cm^{-1}$ ranges of methane using optical frequency comb double-resonance spectroscopy. J Quant Spectrosc Radiat Transfer. 2026;353:109831.
[26] de Martino A, Frey R, Pradere F. Double resonance observation of the ($3\nu_3$, $A_1$) state of methane. Chem Phys Lett. 1984;111:113-6.
[27] Karhu J, Nauta J, Vainio M, Metsälä M, Hoekstra S, Halonen L. Double resonant absorption measurement of acetylene symmetric vibrational states probed with cavity ring down spectroscopy. J Chem Phys. 2016;144.
[28] Hu C-L, Perevalov VI, Cheng C-F, Hua T-P, Liu A-W, Sun YR, et al. Optical–optical double-resonance absorption spectroscopy of molecules with kilohertz accuracy. J Phys Chem Lett. 2020;11:7843-8.
[29] Okubo S, Inaba H, Okuda S, Sasada H. Frequency measurements of the $2\nu_3A_1$-$\nu_3$ band transitions of methane in comb-referenced infrared-infrared double-resonance spectroscopy. Phys Rev A. 2021;103.
[30] Jiang J, McCartt AD. Two-color, intracavity pump-probe, cavity ringdown spectroscopy. J Chem Phys. 2021;155.
[31] Foltynowicz A, Rutkowski L, Silander I, Johansson AC, Silva de Oliveira V, Axner O, et al. Sub-Doppler double-resonance spectroscopy of methane using a frequency comb probe. Phys Rev Lett. 2021;126:063001.
[32] de Oliveira VS, Silander I, Rutkowski L, Soboń G, Axner O, Lehmann KK, et al. Sub-Doppler optical-optical double-resonance spectroscopy using a cavity-enhanced frequency comb probe. Nat Commun. 2024;15:161.
[33] Foltynowicz A, Rutkowski L, Silander I, Johansson AC, Silva de Oliveira V, Axner O, et al. Measurement and assignment of double-resonance transitions to the 8900-9100-$cm^{-1}$ levels of methane. Phys Rev A. 2021;103:022810.
[34] De Oliveira VS, Hjältén A, Silander I, Rosina A, Rey M, Lehmann KK, et al. Combined frequency comb and continuous wave cavity-enhanced optical-optical double-resonance spectrometer in the 1.7 μm range. Opt Express. 2025;33:38776-802.
[35] Lehmann KK, Hjältén A, Silander I, Rey M, Foltynowicz A. Assignment of collision-induced four-level double-resonance transitions in the $3\nu_3 \leftarrow \nu_3$ spectral region of methane. J Chem Phys. 2025;163.
[36] Soboń G, Martynkien T, Mergo P, Rutkowski L, Foltynowicz A. High-power frequency comb source tunable from 2.7 to 4.2 μm based on difference frequency generation pumped by an Yb-doped fiber laser. Opt Lett. 2017;42:1748-51.
[37] Foltynowicz A, Ban T, Masłowski P, Adler F, Ye J. Quantum-noise-limited optical frequency comb spectroscopy. Phys Rev Lett. 2011;107:233002.
[38] Maslowski P, Lee KF, Johansson AC, Khodabakhsh A, Kowzan G, Rutkowski L, et al. Surpassing the path-limited resolution of Fourier-transform spectrometry with frequency combs. Phys Rev A. 2016;93:021802.
[39] Rutkowski L, Masłowski P, Johansson AC, Khodabakhsh A, Foltynowicz A. Optical frequency comb Fourier transform spectroscopy with sub-nominal resolution and precision beyond the Voigt profile. J Quant Spectrosc Radiat Transfer. 2018;204:63-73.

[40] Lehmann KK. Polarization-dependent intensity ratios in double resonance spectroscopy. J Chem Phys. 2023;159.
[41] Gordon IE, Rothman LS, Hargreaves RJ, Hashemi R, Karlovets EV, Skinner FM, et al. The HITRAN2020 molecular spectroscopic database. J Quant Spectrosc Radiat Transfer. 2022;277:107949.
[42] Hjältén A, Silva de Oliveira V, Cao Y, Silander I, Lehmann KK, Foltynowicz A. Optical frequency comb double-resonance spectroscopy of the 9030–9175 $cm^{-1}$ states of ethylene. J Chem Phys. 2026;165.
[43] Foltynowicz A, Masłowski P, Fleisher AJ, Bjork BJ, Ye J. Cavity-enhanced optical frequency comb spectroscopy in the mid-infrared application to trace detection of hydrogen peroxide. Appl Phys B. 2013;110:163-75.
[44] Lyulin OM, Perevalov VI, Morino I, Yokota T, Kumazawa R, Watanabe T. Measurements of self-broadening and self-pressure-induced shift parameters of the methane spectral lines in the 5556–6166$cm^{-1}$ range. J Quant Spectrosc Radiat Transfer. 2011;112:531-9.
[45] de Oliveira VS, Silander I, Sasada H, Okubo S, Inaba H, Lehmann KK, et al. Rotational energy levels in the ground vibrational state of methane with kHz-level accuracy from comb-referenced double-resonance and Lamb-dip spectroscopies. arXiv preprint arXiv:260513060. 2026.
[46] Nikitin AV, Chizhmakova IS, Rey M, Tashkun SA, Kassi S, Mondelain D, et al. Analysis of the absorption spectrum of $^{12}CH_4$ in the region 5855–6250 $cm^{-1}$ of the $2\nu_3$ band. J Quant Spectrosc Radiat Transfer. 2017;203:341-8.
[47] Gordon IE, Rothman LS, Hargreaves RJ, Gomez FM, Bertin T, Hill C, et al. The HITRAN2024 molecular spectroscopic database. J Quant Spectrosc Radiat Transfer. 2026;353:109807.